\documentclass[runningheads,a4paper]{llncs}

\usepackage{amssymb}
\usepackage{graphicx}
\usepackage{nth}
\usepackage{cite}

\usepackage{listings}

\usepackage{url}
\usepackage{epstopdf}
\urldef{\mailsa}\path|stys@jcu.cz|
\newcommand{\keywords}[1]{\par\addvspace\baselineskip
\noindent\keywordname\enspace\ignorespaces#1}

\begin{document}
\lstset{breaklines=true, basicstyle=\footnotesize, tabsize=2}

\mainmatter
\title{Model of the Belousov-Zhabotinsky reaction }
\titlerunning{The model of the B-Z reaction}
\author{Dalibor \v{S}tys \and Tom\'{a}\v{s} N\'{a}hl\'{i}k \and Anna Zhyrova \and Renata Rycht\'{a}rikov\'{a} \and \v{S}t\v{e}p\'{a}n Pap\'{a}\v{c}ek \and Petr C\'{i}sa\v{r}}

\authorrunning{\v{S}tys et al.}

\institute{University of South Bohemia in \v{C}esk\'{e} Bud\v{e}jovice,\\ Faculty of Fisheries and Protection of Waters,\\ South Bohemian Research Center of Aquaculture and Biodiversity of Hydrocenoses,\\ Institute of Complex Systems,\\ Z\'{a}mek 136, 373 33 Nov\'{e} Hrady, Czech Republic
\mailsa\\
\url{http://www.frov.jcu.cz/cs/ustav-komplexnich-systemu-uks}}

\toctitle{Lecture Notes in Bioinformatics}
\tocauthor{Dalibor \v{S}tys et al.}
\maketitle

\begin{abstract}
The article describes results of the modified model of the Belousov-Zhabotinsky reaction which resembles well the limit set observed in an experiment in the Petri dish. We discuss the concept of the ignition of circular waves and show that only an asymmetrical ignition leads to the formation of spiral structures. From the qualitative assumptions on the behavior of dynamic systems we conclude that the reactants in the Belousov-Zhabotinsky reaction likely forms a regular grid. 

\end{abstract}

\section{Introduction}

The Belousov-Zhabotinsky (B-Z) reaction \cite{Belousov1959, Zhabotinsky1964, Zhabotinsky1964b} is an experimentally easily accessible example of chemical self-organization. Many chemical reactions which contribute to this process are already known, e.g. \cite{RovinskyandZhabotinsky1988}, and new chemical reactions and compounds are being added every year (e.g., \cite{Hagelsteinal2013}).

Until recently, the reaction-diffusion simulations based on PDEs, which are central models of chemical processes, have been used extensively to explain the self-organizing Belousov-Zhabotinsky reaction \cite{VanagandEpstein2009}. This kind of simulation expects an instantaneous chemical change. However, in reality, any chemical reaction is a combination of the physical breaking of chemical bonds, splitting, and diffusion of new chemical moieties over a time span -- a defined elementary timestep needed for the progression of the spatial-limited chemical process. Therefore, for the description of the B-Z reaction, we chose a kind of cellular automaton -- a hodgepodge machine, see e.g. \cite{Gerhardt1989,Dewdney1988} -- for the time-spatially discrete simulation.

The hodgepodge machine is able to correctly simulate the final state of the reaction course of the B-Z reaction as well as the formation of the mixture of spirals and waves. According to Wuensche \cite{Wuensche2011}, the final state is considered to be the limit set and the course of the experiment is the trajectory through the state space. In this way, all events which precede the establishment of spirals and waves are state space trajectories by which the system travels through ro reach the ergodically behaving dynamic process of the limit set.

\medskip
\noindent{\it Hodgepodge machine NetLogo algorithm -- standard steps}~\cite{Wilensky2003}
\begin{lstlisting}[numbers=left]
;; all calculations are made in int

const
     maxState; the maximal state level 
     k1 ;; the weight of the cell in the ignition state from the interval (0, maxState)
     k2 ;; the weight of the cell with ignition state maxState 
     g ;; the number of levels crossed in one simulation step
 
patch-own
     state_{x,y}[t] ;; the state level of the patch in step t
     a ;; the number of the cells in the Moore neighborhood in the state from the interval (0, maxState)
     b ;; the number of cells in the Moore neighborhood in maxState 
	 
to setup
	foreach x,y 
	state_{x,y} = random(maxState + 1)
end

begin 
	foreach x,y
	count a
	count b
	ifelse state_{x,y}[t] = maxState
	state_{x,y}[t+1] = 0
		ifelse state_{x,y}[t] = 0
		state_{x,y}[t+1] = int(a/k1) + int(b/k2)
			state_{x,y}[t+1] = int((sum [state] of neighbors)/(a+b+1))+g)
				if state_{x,y}[t+1] > maxState
        			state_{x,y}[t+1] = maxState
end	
\end{lstlisting}

\medskip

In brief, we consider four processes in the hodgepodge machine algorithm (see above) which describe the behavior of the chemical reaction at the quantum (electron) scale:
\begin{enumerate}
\item Deexcitation -- phase transition -- bond breakage upon reaching the maximal level (lines 23--24 and 28--29),
\item spreading the "infection" from neighboring cells (lines 25--26),
\item excitation process which is simulated by the acquisition of cell levels (energy transfer) from the average of neighboring cells (line 27), and
\item growth of the excitation inside the cell (line 27).
\end{enumerate}
Items 1--2 are the most speculative processes, but we have at least the experimental evidence for them. Further, we assume that processes 3 and 4 occur on the border and inside of the structural element, respectively. In this aspect, the hodgepodge machine may be considered as the most elementary example with (a) only forward (excitation) processes and (b) a linear increase of the process inside the spatial element and on its border. The automaton depends upon 4 parameters, $maxState$, $k_1$, $k_2$, $g$,  as well as on initial (an ignition phase) and boundary conditions (usually periodic).

In this article we report some experimental modifications on the hodgepodge machine. We started with only a few ignition points and varied the number of states and the constants $g$ and $maxState$. Constants $k_{1}$ and $k_{2}$ are critical for the course of the system dynamics. If $\frac{1}{k_{1}} + \frac{0}{k_{2}} < 0.5$ and $\frac{0}{k_{1}} + \frac{1}{k_{2}} < 0.5$, we observed a limit set of waves and spirals. In all other cases, we observed filaments. We examine the influence of constants $k_{1}$, $k_{2}$, $g$, and $maxState$.

\section{Methods}

We used the Wilensky NetLogo modeling system \cite{Wilensky2003} for its versatility and, in the discussion, Wuensche's terminology \cite{Wuensche2011}. Our proposed modification of the Wilensky implementation is as follows:

\medskip
\noindent
{\it Hodgepodge machine NetLogo algorithm -- exponential start}~\cite{url1}
\begin{lstlisting}
  ..........
  
const
     ......
     
     meanPosition
     ;; the mean of the random exponential distribution of the starting state (a new input constant)
     
     .......... 
	 
setup
	foreach x, y 
	state_{x,y} = random-exponential((maxState + 1) * meanPosition)                          ;; modified l. 16
  end
..............
      
begin 
	foreach x,y

	.........
	
		state_{x,y}[t+1] = round(a/k1 + b/k2)    ;; modified l. 26
		
	...........

end
\end{lstlisting}

\medskip
As shown, the Wilensky model was modified in two ways. The first change (line 16) was introduced to ensure that the simulation started with a small number of ignition points, similar to those found in the experiment. By a proper set-up of the constants $maxState$ and $meanPosition$, we achieved the condition in which the majority of points are in the initial $state_{x,y}$ higher than $maxState$. In the next step, we obtained a very few points in the non-zero state. This modification allowed us to observe the early phase of the wave formation in a way different from that of the original hodgepodge machine, but more similar to the evolution observed in the experiment. 

\begin{figure}
\centering
\includegraphics[width=1\textwidth]{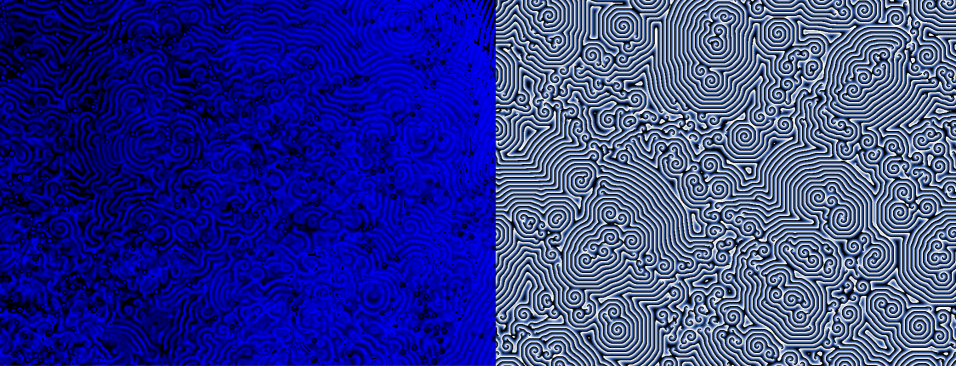}
\caption{Comparison of the Belousov-Zhabotinsky experiment record in the blue camera channel (left) and its NetLogo simulation (right).}
\label{fig:BZ-model_comparison}
\end{figure}

The second modification (line 26) was implemented in order to allow us to start from these few ignition points. However, the non-intentional result of this modification was the increased similarity between the experiment and its simulation (model) (Fig. \ref{fig:BZ-model_comparison}), which illustrates the importance of this ignition rule.

The calculation was performed on a canvas of 1000 $\times$ 1000 cells. With a finer grid, the non-idealities of the periodic boundary did not suppress the system's behavior. Namely, the calculation with only a few points of ignition did not lead to the formation of regular structures.

In this paper, results achieved at $maxState$ = 200 and $g$ = 28 are discussed. Their role in the model is described in the next article in this volume. 

For comparison to the proposed simulation, the B-Z reaction was performed as published previously \cite{ZhyrovaStys2014}.

\section{Results}
\subsection{Traveling through the basin of attraction}

In the B-Z reaction, instead of observing the formation of a stable Turing pattern \cite{Turing1955}, we saw a regular stepwise interchange of self-similar structures which are in fact cellular automata, i.e., discrete dynamic networks \cite{Wuensche2011}. In these systems, Gardens of Eden -- configurations which the system may not reach from any previous, intermediate configurations -- are sought. The next step in the trajectory consists of configurations that have a parental configuration. Many different Gardens of Eden often lead to one common structure, and there may be a few such ``common" configurations on the trajectory towards the limit set. It is the state that a dynamical system reaches after an infinite amount of time. The limit set is where the system exhibits ergodic behavior. In discrete dynamic networks, it is either a fixed point or a periodically interchanging set of configurations. 

In a continuous system in the two-dimensional space (e.g., plane), the properties of the limit set are governed by the Poincare-Bendixson Theorem \cite{Bendixson1901}. It states that if a differentiable real dynamical system is defined on an open subset of the (two-dimensional) plane, then every non-empty compact $\omega$-limit set of an orbit is either a fixed point, a periodic orbit, or a set composed of a finite number of fixed points connected by homoclinic and heteroclinic orbits. In higher dimensions, there may be other types of behavior, such as chaotic behavior or strange attractor. In discrete dynamic systems, chaotic behavior can arise in two or even one-dimensional systems. 

In the final state of both the hodgepodge machine simulation and the B-Z experiment (Fig. 1), we observed a state which can be characterized as complex, i.e., having medium entropy and high variance. It is unclear how we can apply this terminology to the realm of multilevel cellular automata with more complicated rules, because the field is much less developed. However, we can ensure that it is not any type of periodic behavior predicted by the Poincare-Bendixson Theorem, because discrete systems typically display spirals \cite{Fischetal1991}.

\subsection{Influence of the $k_1$ and $k_2$ variation}

In this section, results achieved at $maxState$ = 200 and $g$ = 28 are discussed. The role of $maxState$ and $g$ ratio is described later in the text. 

\begin{figure}
\centering
\includegraphics[width=1\textwidth]{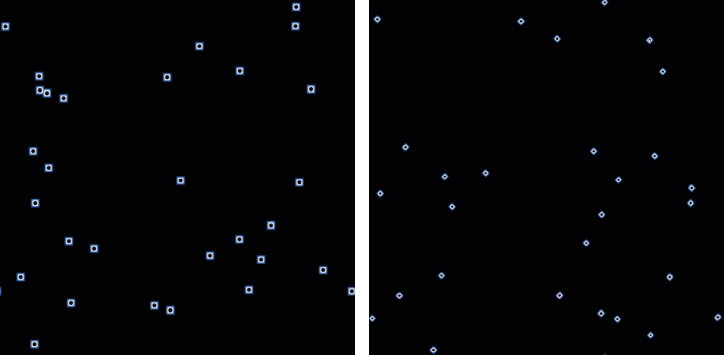}
\caption{The modified hodgepodge simulation in step 15 for $k_{1}=2$ and $k_{2}=2$ (left) and $k_{1}=3$ and $k_{2}=3$ (right).}
\label{fig:Trajectory_15}
\end{figure}

\begin{figure}
\centering
\includegraphics[width=1\textwidth]{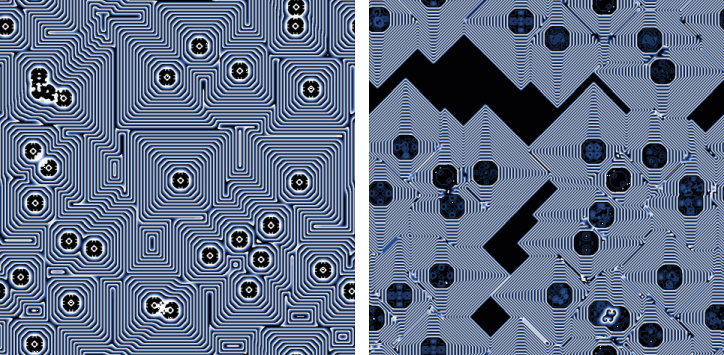}
\caption{The modified hodgepodge simulation in step 200 for $k_{1}=2$ and $k_{2}=2$ (left) and $k_{1}=3$ and $k_{2}=3$ (right).}
\label{fig:Trajectory_200}
\end{figure}

\begin{figure}
\centering
\includegraphics[width=1\textwidth]{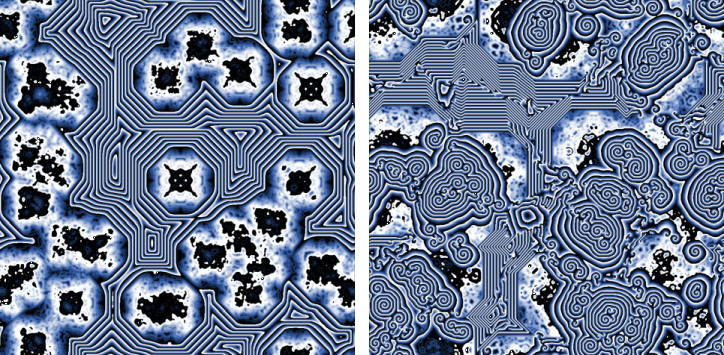}
\caption{The modified hodgepodge simulation in step 700 for $k_{1}=2$ and $k_{2}=2$ (left) and $k_{1}=3$ and $k_{2}=3$ (right).}
\label{fig:Trajectory_700}
\end{figure}

\begin{figure}
\centering
\includegraphics[width=1\textwidth]{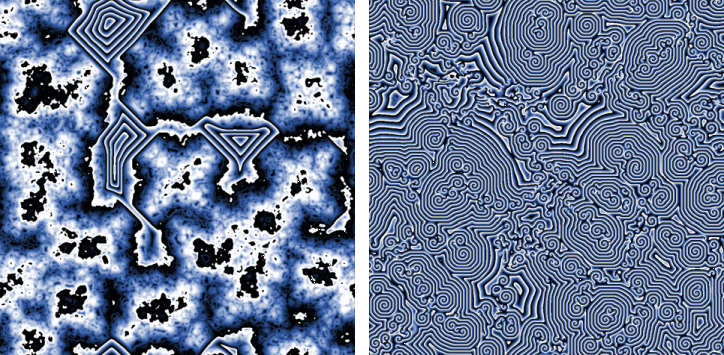}
\caption{The modified hodgepodge simulation in step 1200 for $k_{1}=2$ and $k_{2}=2$ (left) and $k_{1}=3$ and $k_{2}=3$ (right).}
\label{fig:Trajectory_1200}
\end{figure}

\begin{figure}
\centering
\includegraphics[width=1\textwidth]{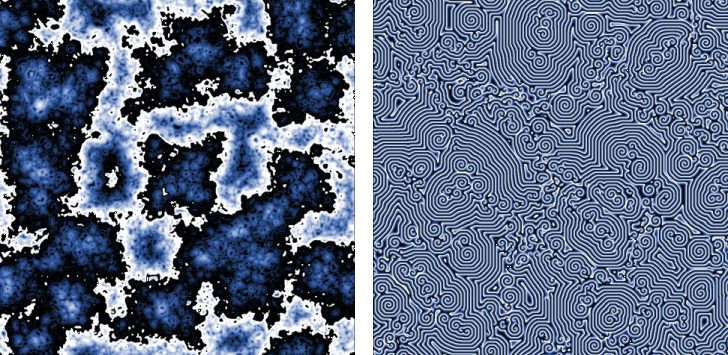}
\caption{The modified hodgepodge simulation in step 2000 for $k_{1}=2$ and $k_{2}=2$ (left) and $k_{1}=3$ and $k_{2}=3$ (right).}
\label{fig:Trajectory_2000}
\end{figure}

Figs. \ref{fig:Trajectory_15}--\ref{fig:Trajectory_2000} show two possible trajectories of the hodgepodge machine to the basin of attraction. When $k_{1}<3$ or $k_{2}<3$, the initial concentric waves are circular. It is enabled by the production of waves from a single point, because $\mbox{round}(\frac{1}{2} + \frac{0}{2}) > 0$, $\mbox{round}(\frac{1}{2} + \frac{1}{2}) > 0$, and $\mbox{round}(\frac{0}{2} + \frac{1}{2}) > 0$. In contrast, if $k_{1}\leq 3$ or $k_{2}\leq 3$, we need at least one non-zero cell in the neighborhood of the zero cell to achieve the same effect, e.g., $\mbox{round}(\frac{1}{3} + \frac{1}{3}) > 0$. This subtlety has a very strong influence on the system's behavior.

In the early phase of the simulation (Fig. \ref{fig:Trajectory_15}),  for $k_{1}=2$ and $k_{2}=2$, we observed evolution of squares whose center appears to be circular. In contrast, the case $k_{1}=3$ and $k_{2}=3$ produced octagons with four long and four short sides. 

Both cases of step 200 (Fig. \ref{fig:Trajectory_200}) created a system of evolved waves. We observed a second type of the structure in the centre of the travelling wave system. For $k_{1}=3$ and $k_{2}=3$, we already observed a spiral doublet. 

In step 700 (Fig. \ref{fig:Trajectory_700}), different $k$ values continued to exhibit different behaviors. Spirals with a new type of wave emanating from the centre were ``more stable" than the filamentous structures that appeared in both cases in Fig. \ref{fig:Trajectory_200}).

In step 1200 (Fig. \ref{fig:Trajectory_1200}), the early wave structures vanished. For $k_{1}=2$ and $k_{2}=2$, the image was full of thin filamentous structures, whereas for $k_{1}=3$ and $k_{2}=3$, the image was almost identical to the final structure of the limit set. 

Finally, in step 2000 (Fig. \ref{fig:Trajectory_2000}), for $k_{1}=3$ and $k_{2}=3$, the simulated structure fully matched the experiment structure. The ratio of spirals, wave structure fragments, and intermediate objects was stable within a certain limit. The case $k_{1}=2$ and $k_{2}=2$ was full of interchanging filamentous structures, which would, in further simulation, undergo a slow broadening into homogeneity (never observed in simulations). It seems that the limit set in the latter case is a certain type of coexistence of broad filaments. Since we do not have any experimental parallel to the filamentous type of structures, we continue with the analysis of the spiral structure.

\subsection{Influence of the maximal number of cell states $maxState$ on the limit sets}

\begin{figure}
\centering
\includegraphics[width=0.6\textwidth]{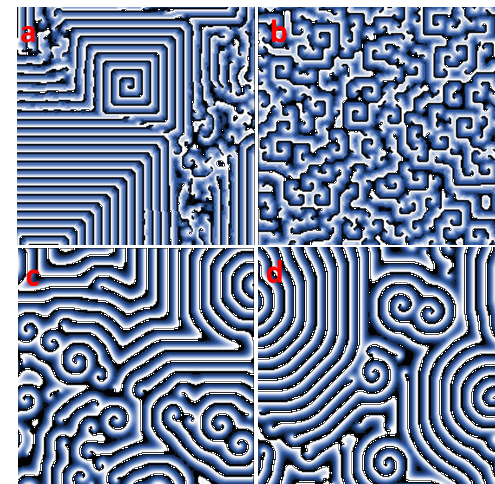}
\caption{The influence of the number of cell states $maxState$ on the limit sets at similar $g/maxState$ ratios of 1:7 (\textbf{a}), 3:20 (\textbf{b}), 28:200 (\textbf{c}), and 280:2000 (\textbf{d}). Total overviews of the limit sets' structures.}
\label{fig:Number_of_levels}
\end{figure}

Multilevel cellular automata are often analyzed for at most 8 cell states, e.g. \cite{Wuensche2011}. Here, we tested the modified Wilensky hodgepodge machine algorithm by changing the maximal number of the achievable cell states $maxStates$ (Fig. \ref{fig:Number_of_levels}).

\begin{figure}
\centering
\includegraphics[width=0.6\textwidth]{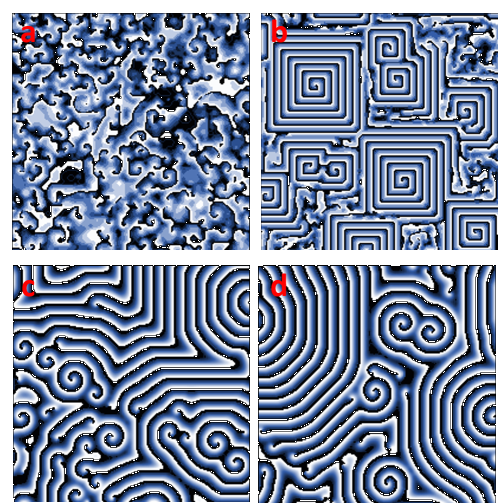}
\caption{The influence of the number of cell states $maxState$ on limit sets at similar $g/maxState$ ratios of 1:6 (\textbf{a}), 1:7 (\textbf{b}), 28:200 (\textbf{c}), and 280:2000 (\textbf{d}). Detailed views of the limit sets' structures.}
\label{fig:Number_of_levels_cut}
\end{figure}

The limit set -- a mixture of waves and spirals -- highly similar to those observed in the B-Z experiment was achieved at the $g/maxState$ ratio of 28:200 (compare Figs. \ref{fig:Number_of_levels}\textbf{c} and \ref{fig:Number_of_levels_cut}\textbf{c} with Fig. \ref{fig:Experimental_trajectory} in 210 s). We observed branched spirals there (Fig. \ref{fig:Number_of_levels}\textbf{c}). At the same ratio and higher $maxState$ value (Figs. \ref{fig:Number_of_levels}\textbf{d}), the spirals were even more smooth. To obtain these results, we kept the ratio constant and simplified it to 7:50. The simulation at the $g/maxState$ ratio of 3:20 brought up the limit set, which was also similar to the chemical experiment. The $g/maxState$ ratio of 1:7 (Fig. \ref{fig:Number_of_levels}\textbf{a}) led to the formation of square spirals reported earlier \cite{Wuensche2011, Fischetal1991}. The next ratio 28:200 (close to 1:7) gave results most similar to the experiment. The $g/maxState$ ratio of 1:8 (Fig. \ref{fig:Number_of_levels}\textbf{b}) already showed spirals and waves like the B-Z experiment. For all comonnly studied "small-number" ratios, the simulation and reaction in the limit sets differ. At $maxState$ $<$ 20 the spiral evolutions were already qualitatively similar.

Among the low-number $maxState$ levels, the simulation at the 1:7 ratio was extraordinary due to slow evolution to the limit set. Reaching the limit set took much higher number of simulation steps (37,000) than in other cases (e.g., 2,500 steps at $g/maxState$ = 1:8). Indeed, 8 levels, which correspond to the number of pixels in the Moore neighborhood, already tend to make octagons (read below). At one level below (at 7 levels), the system's behavior was already changed. However, further research needs to be done to understand this phenomenon.

\subsection{Temporarily organized structures in the initial phases and the limit sets at different $g/maxState$ ratios}

\begin{figure}
\centering
\includegraphics[width=0.85\textwidth]{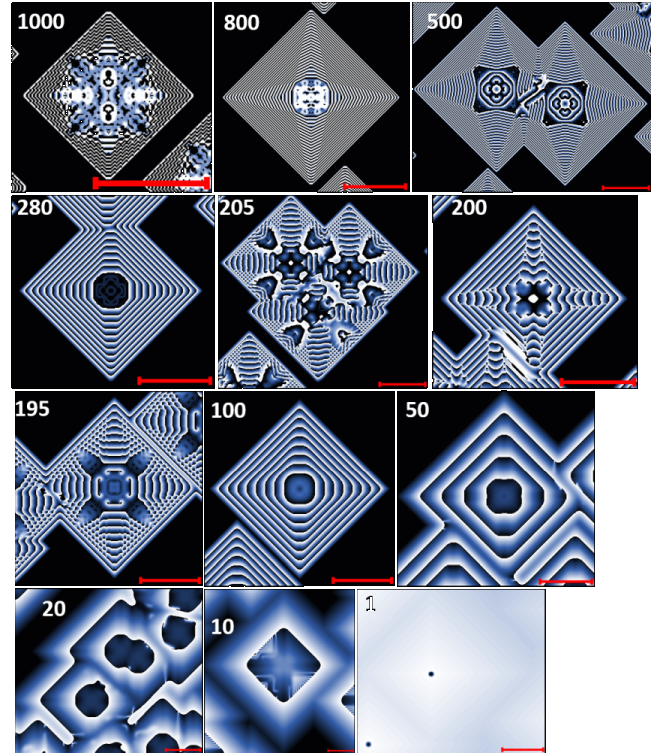}
\caption{The initial states for selected values of $g$ at the constant $maxState$ = 2000. Each of the scale bars corresponds to 100 px. As in the previous figures, the black color corresponds to the state level 0, white color to the maximal state value (set by the constant $maxState$).}
\label{fig:typology_of_central_structures}
\end{figure}

\begin{figure}
\centering
\includegraphics[width=0.85\textwidth]{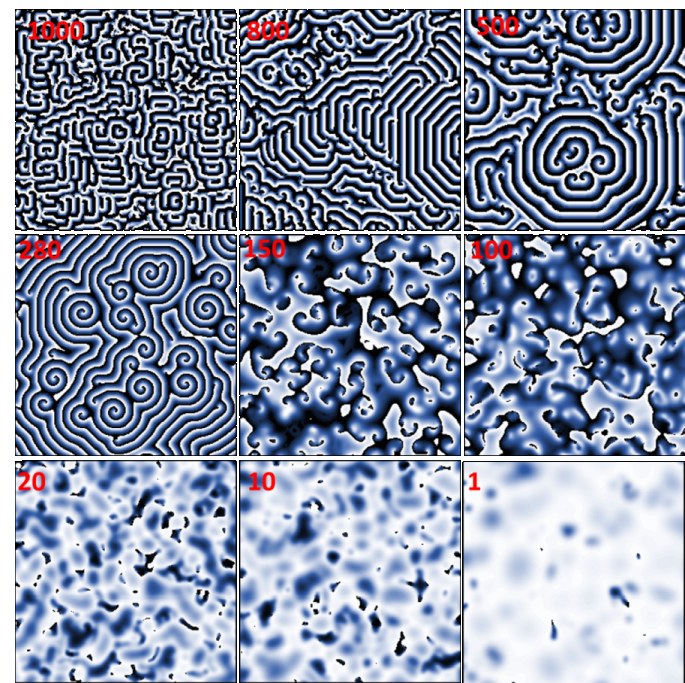}
\caption{The limit sets for selected $g$ at $maxState$ = 2000. From the upper left to lower right image, the values of $g$ are  1000, 800, 500, 280, 150, 100, 20, 10, and 1, respectively.}
\label{fig:typology_of_limit_sets}
\end{figure}

The temporarily organized structures on the state-space trajectory of the simulation are those on one of the paths of the discrete dynamic network \cite{Wuensche2011}. In Figs. \ref{fig:typology_of_central_structures} and \ref{fig:typology_of_limit_sets} we show the shape dependence of the temporarily organized structures in the initial phases and limit sets on the $g$ value at the constant $maxState$ value (i.e., on the decreasing $g/maxState$ ratio).

Various ignition points, which are assumed to be Gardens of Eden, gave octagonal structures (Fig. \ref{fig:typology_of_central_structures}), whose interiors were typical for the given Gardens of Eden. At the high $g/maxState$ ratio, after the passage of early waves when the interior octagon became regular, a centrally symmetrical structures appeared in the centre. The shapes of such initial central structures are dependent on the value of the $g/maxState$ ratio. We do not describe here the mechanism how these structures arise.

The early square waves gradually broadened with decreasing\linebreak$g/maxState$ ratio (Fig. \ref{fig:typology_of_central_structures}). The lower the $g/maxState$, the more diffuse the structures appeared. At $g/maxState \leq$ 50:2000 we obtained a diffusive central object surrounded by the spreading wavefronts. With $g/maxState$ of 10:2000, the fuzzy diffusive wavefronts were becoming more sparse. At $g/maxState =$ 1:2000 the second wave stopped evolving and the circular dark central structure appeared inside the dense diffusion.

Thus, the $g/maxState$ ratio can explain the thickness of the early waves and determine the number of states achieved in one time interval. This suggests that the wave thickness and density can, to some extent, explain other phenomena such as the built-in local asymmetry of the space, which ignites all additional phenomena and determines the state space trajectory of the process to its limit sets.

The limit sets (Fig. \ref{fig:typology_of_limit_sets}) first emerge from a fully spiral form at $g/maxState$ = 1000:2000. As $g/maxState$ decrease down to 280:2000, the spirals more and more resemble those found in a natural chemical reaction. The waves were broadening and by $g/maxState$ = 150:2000 mature spirals could no longer be observed. Later, the organized structures were broken into a diffusive mixture of darker and lighter stains. At the limit $g/maxState \rightarrow 0$, we are likely to obtained a uniform intensity. It shows the unsuitability of the reaction-diffusion model for the description of the B-Z reaction \cite{VanagandEpstein2009}.

\subsection{Evolution of the state space trajectory}

\begin{figure}
\centering
\includegraphics[width=0.85\textwidth]{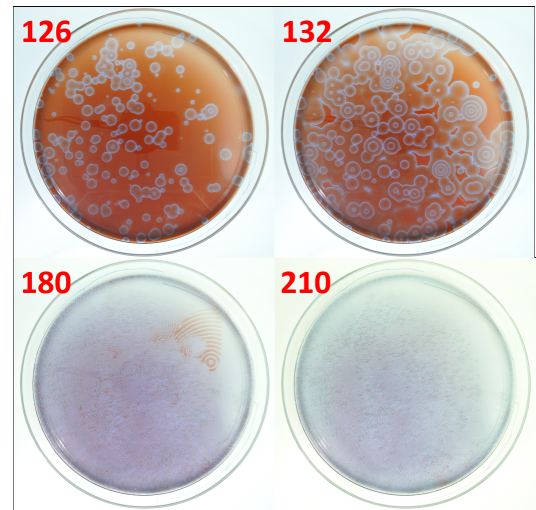}
\caption{Segments of the experimentally determined trajectory of the Belousov-Zhabotinsky reaction. Numbers correspond to the time interval (in seconds) from the initial mixing of the reactants.}
\label{fig:Experimental_trajectory}
\end{figure}

The main goal of the hodgepodge machine simulation is to test its consistency with the course of the B-Z reaction (an experimental trajectory). The segments of the reaction are depicted in Fig. \ref{fig:Experimental_trajectory}.

\begin{figure}
\centering
\includegraphics[width=0.85\textwidth]{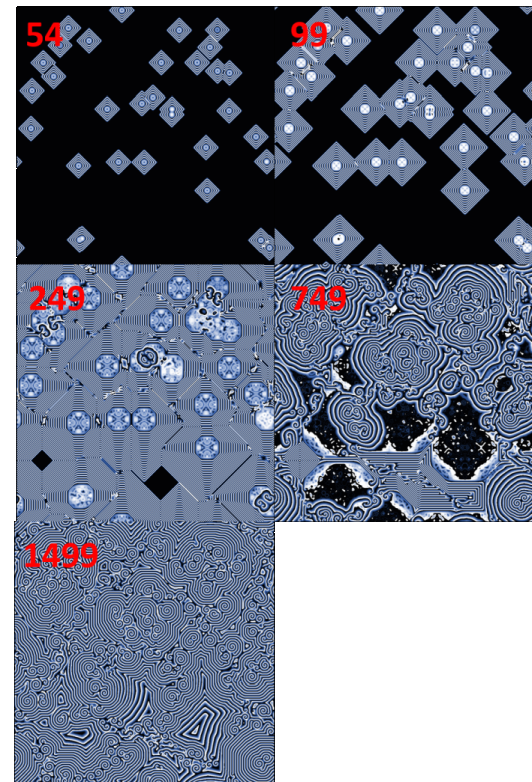}
\caption{The phases of the state-space trajectory at $g/maxState = 280:2000$. \textbf{Step 54} -- The octagonal waves in the centre collapses and is replaced by the fractal structure. \textbf{Step 99} -- In the places of the collision of two central structures, early spiral structures arise. \textbf{Step 249} -- Spiral structures are being surrounded by the elliptical wavefront. \textbf{Step 749} -- The simulation canvas is being filled by new spiral waves. \textbf{Step 1499} -- Structures become almost regular by "compression" of waves of various origins.}
\label{fig:Typical_state_trajectory_phases}
\end{figure}

Despite the similarity of the limit set between the simulation at $g/maxState$ ratio = 280:2000 (Figs. \ref{fig:Number_of_levels}\textbf{c} and \ref{fig:Number_of_levels_cut}\textbf{c}) and the chemical experiment, the trajectories towards these limit sets were a bit different. In the case of the simulation (Fig. \ref{fig:Typical_state_trajectory_phases}), after the ignition, the structures evolved in a sequence of square waves with round centers. After that, the first spiral doublet arose in the vicinity of two central objects such that the waves collided in one point. The doublet soon became surrounded by an elliptical wave thicker than the early square wave. This elliptical wave system had an expanding "diffusive" character until it was compressed by another system of elliptical waves which slowly evolved at other points.

In case of the model, early circular waves analogous to those in the experiment, were achieved only at very low $g/maxState$ ratios, when the waves were broad and the spirals were not formed (e.g., Fig. \ref{fig:typology_of_central_structures} for $g = 20$). This suggests that in the early phase of the experimental trajectory, there is a second process which brings the evolution of a different trajectory. Later, when this process is exhausted, the evolution of the system follows the path described by the hodgepodge machine model.

\section{Discussion}

The formation of spirals is somewhat mysterious. There currently does not exist a mathematical equation which describes the ``robustness" or ``stability" of the spiral. We also do not currently have a mathematical equation parallel to the Poincare-Bendixon Theorem for differentiable systems to describe our discrete system. Although, if it is a rule for discrete systems, why does it appear in the chemical systems in which we assume differentiability?  

From the chemical point of view, the fundamental question in the interpretation of the B-Z reaction is whether the observed pattern is a result of a continuous behavior or whether the space is somehow discretised. Stable systems with two liquid phases -- emulsions -- are observed in many natural examples.

We propose that the similarity between the final phase of the experiment and the limit set of the simulation (Fig. \ref{fig:BZ-model_comparison}) is not a coincidence. This suggests that the experimental problem may be discussed in the same way as the simulation. For the continuous case in the plane, where the Poincare-Bendixson Theorem holds, we should observe a fixed point, a limit cycle or a set of homoclinic and heteroclinic orbitals. However, this is not what we observed. Instead, we saw a mixture of wave fragments and spirals, where the spirals seem to be confined to discrete spatial arrangement. We conclude that, beyond the reasonable doubts, the structures observed in the B-Z reaction are the consequences of a certain kind of the space discretisation.

The fact that spirals are observed only in the case where two non-zero pixels exist in the neighborhood indicates that there must be a certain kind of local spatial asymmetry needed for spiral formation. Detailed discussion of this phenomenon is rather extensive and will be addressed later.

The time-spatial discretization is the idealized process, where the time element determines (a) a set of events achieved at the time within a spatial element (a cell) -- the $g$ constant -- and (b) the rest of events on the spatial element's (cell's) borders -- the $maxState$ constant. We may consider analogies for such behaviour in known energy transfer processes, i.e. the difference between resonance energy transfer and energy transfer in systems of overlapping orbitals.

The analysis of these constants showed that the $g/maxState$ ratio determined the thickness of waves and the course of the state space trajectory irrespective to the total $maxState$ value. The decreasing $g/maxState$ ratio led to the loss of the wave structure, which eventually resulted in a fully diffusive picture. There also exists a lower limit of $maxState$ for the appearance successful course of the simulation. The higher the $maxState$ value, the smoother the spirals were.

\subsubsection*{Acknowledgments.} 
This work was financially supported by CENAKVA (No. CZ.1.05/2.1.00/01.0024), CENAKVA II (No.LO1205 under the NPU I program) and The CENAKVA Centre Development (No. CZ.1.05/2.1.00/19.0380). Authors thank to Petr Jizba and Jaroslav Hlinka for important discussions and to Kaijia Tian for edits.

\end{document}